\documentclass[runningheads]{llncs}
\usepackage{microtype}
\usepackage{graphicx}
\usepackage{wrapfig}
\usepackage{sidecap}

\usepackage{booktabs} 
\usepackage{makecell}

\usepackage{amsmath}
\usepackage{amssymb}
\usepackage{mathtools}
\usepackage{dsfont}
\usepackage{tabularx}

\input{Qcircuit}
\usepackage{physics}
\usepackage{bm}
\usepackage{bbm}
\usepackage{tikz}

\usepackage{underscore}
\usepackage[export]{adjustbox}

\usepackage{subfig}

\definecolor{jancolor}{rgb}{0.4,0.6,0.2}

\newcommand{\params}{\ensuremath{\bm{\theta}}}

\newcommand{\CZ}[1]{%
\begin{tikzpicture}[#1]%
\draw[thick] (0ex,-0.5ex) -- (0ex, 0.5ex);%
\filldraw [black] (0ex, -0.5ex) circle (0.7pt);
\filldraw [black] (0ex, 0.5ex) circle (0.7pt);
\end{tikzpicture}%
}
\newcommand{\vtheta}{\boldsymbol{\theta}}
\newcommand{\vTheta}{\boldsymbol{\Theta}}

\begin{document}
\title{Hyperparameter Importance of Quantum Neural Networks Across Small Datasets}
\titlerunning{Hyperparameter Importance of quantum neural networks}

\author{Charles Moussa\orcidID{0000-0002-5387-564X} \and
Jan N. van Rijn\orcidID{0000-0003-2898-2168} \and
Thomas B\"{a}ck\orcidID{0000-0001-6768-1478} \and
Vedran Dunjko\orcidID{0000-0002-2632-7955}}
\authorrunning{C. Moussa et al.}

\institute{LIACS, Leiden University, Niels Bohrweg 1, 2333 CA Leiden, Netherlands}
\maketitle              
\begin{abstract}
As restricted quantum computers are slowly becoming a reality, the search for meaningful first applications intensifies. In this domain, one of the more investigated approaches is the use of a special type of quantum circuit - a so-called quantum neural network -- to serve as a basis for a machine learning model. Roughly speaking, as the name suggests, a quantum neural network can play a similar role to a neural network. However, specifically for applications in machine learning contexts, very little is known about suitable circuit architectures, or model hyperparameters one should use to achieve good learning performance. In this work, we apply the functional ANOVA framework to quantum neural networks to analyze which of the hyperparameters were most influential for their predictive performance. We analyze one of the most typically used quantum neural network architectures. We then apply this to $7$ open-source datasets from the OpenML-CC18 classification benchmark whose number of features is small enough to fit on quantum hardware with less than $20$ qubits. Three main levels of importance were detected from the ranking of hyperparameters obtained with functional ANOVA. Our experiment both confirmed expected patterns and revealed new insights. For instance, setting well the learning rate is deemed the most critical hyperparameter in terms of marginal contribution on all datasets, whereas the particular choice of entangling gates used is considered the least important except on one dataset. This work introduces new methodologies to study quantum machine learning models and provides new insights toward quantum model selection.

\keywords{Hyperparameter importance \and Quantum Neural Networks \and Quantum Machine Learning.}
\end{abstract}

\section{Introduction}
\label{intro}
Quantum computers have the capacity
to efficiently solve computational problems believed to be intractable for classical computers, such as factoring~\cite{Shor1999PolynomialTimeAF} or simulating quantum systems~\cite{qsim}. However, with the Noisy Intermediate-Scale Quantum era~\cite{Preskill2018quantumcomputingin}, quantum algorithms are confronted with many limitations (e.g., the number of qubits, decoherence, etc). Consequently, hybrid quantum-classical algorithms were designed to work around some of these constraints while targeting practical applications such as chemistry~\cite{VQE}, combinatorial optimization~\cite{QAOA}, and machine learning~\cite{PQC}. Quantum models can exhibit clear potential in special datasets where we have theoretically provable separations with classical models~\cite{robustseparation,qrl1,chemistryqml,Swekegen}. More theoretical works also study these models from a generalization perspective~\cite{genboundpqc}. 
Quantum circuits with adjustable parameters, also called quantum neural networks, have been used to tackle regression~\cite{qcl}, classification~\cite{qsvm}, generative adversarial learning~\cite{qgan}, and reinforcement learning tasks~\cite{qrl1,qrl2}.
\par
However, the value of quantum machine learning on real-world datasets is
still to be investigated in any larger-scale systematic fashion~\cite{largeqml,largeqmlgoogle}. Currently, common practices from machine learning, such as large-scale benchmarking, hyperparameter importance, and analysis have been challenging tools to use in the quantum community~\cite{schuldperspective}. Given that there exist many ways to design quantum circuits for machine learning tasks, this gives rise to a hyperparameter optimization problem. However, there is currently limited intuition as to which hyperparameters are important to optimize and which are not. Such insights can lead to much more efficient hyperparameter optimization~\cite{brazdil2022metalearning,feurerarxiv20a,mohr2022learning}.
\par
In order to fill this gap, we employ functional ANOVA~\cite{Hutter2014,Sobol1993}, a tool for assessing hyperparameter importance. This follows the methodology of~\cite{hpimp,resnetfanova}, who employed this across datasets, allowing for more general results. For this, we selected a subset of several low-dimensional datasets from the OpenML-CC18 benchmark~\cite{openmlcc18}, that are matching the current scale of simulations of quantum hardware. We defined a configuration space consisting of ten hyperparameters from an aggregation of quantum computing literature and software. We extend this methodology by an important additional verification step, where we verify the performance of the internal surrogate models. Finally, we perform an extensive experiment to verify whether our conclusions hold in practice. While our main findings are in line with previous intuition on a few hyperparameters and the verification experiments, we also discovered new insights. For instance, setting well the learning rate is deemed the most critical hyperparameter in terms of marginal contribution on all datasets, whereas the particular choice of entangling gates used is considered the least important except on one dataset.

\section{Background}
\label{background}

In this section, we introduce the necessary background on functional ANOVA, quantum computing, and quantum circuits with adjustable parameters for supervised learning.

\subsection{Functional ANOVA}
When applying a new machine learning algorithm, it is unknown which hyperparameters to modify in order to get high performances on a task. Several techniques for hyperparameter importance exist, such as functional ANOVA~\cite{fanova}. The latter framework can detect the importance of both individual hyperparameters and interaction effects between different subsets of hyperparameters. We first introduce the relevant notations following, based upon the work by~\cite{Hutter2014}. 
\par
Let $A$ be an machine learning algorithm that has $n$ hyperparameters with domains $\Theta_1, \ldots, \Theta_n$ and \emph{configuration space} $\vTheta = \Theta_1 \times \ldots \times \Theta_n$. An instantiation of $A$ is a vector $\vtheta = \{ \theta_1, \ldots, \theta_n \}$ with $\theta_i \in \Theta_i$ (this is also called a \emph{configuration} of $A$). A partial instantiation of $A$ is a vector $\vtheta_U = \{ \theta_{i_1}, \ldots, \theta_{i_k} \}$ with a subset $U=\{i_1, \ldots, i_k\} \subseteq N = [n] = \{1, \dots, n\}$ of the hyperparameters fixed, and the values for other hyperparameters unspecified. Note that $\vtheta_N = \vtheta$. 
\par
\begin{figure}[!tb]
\centering
\subfloat[]{\includegraphics[width=0.33\textwidth]{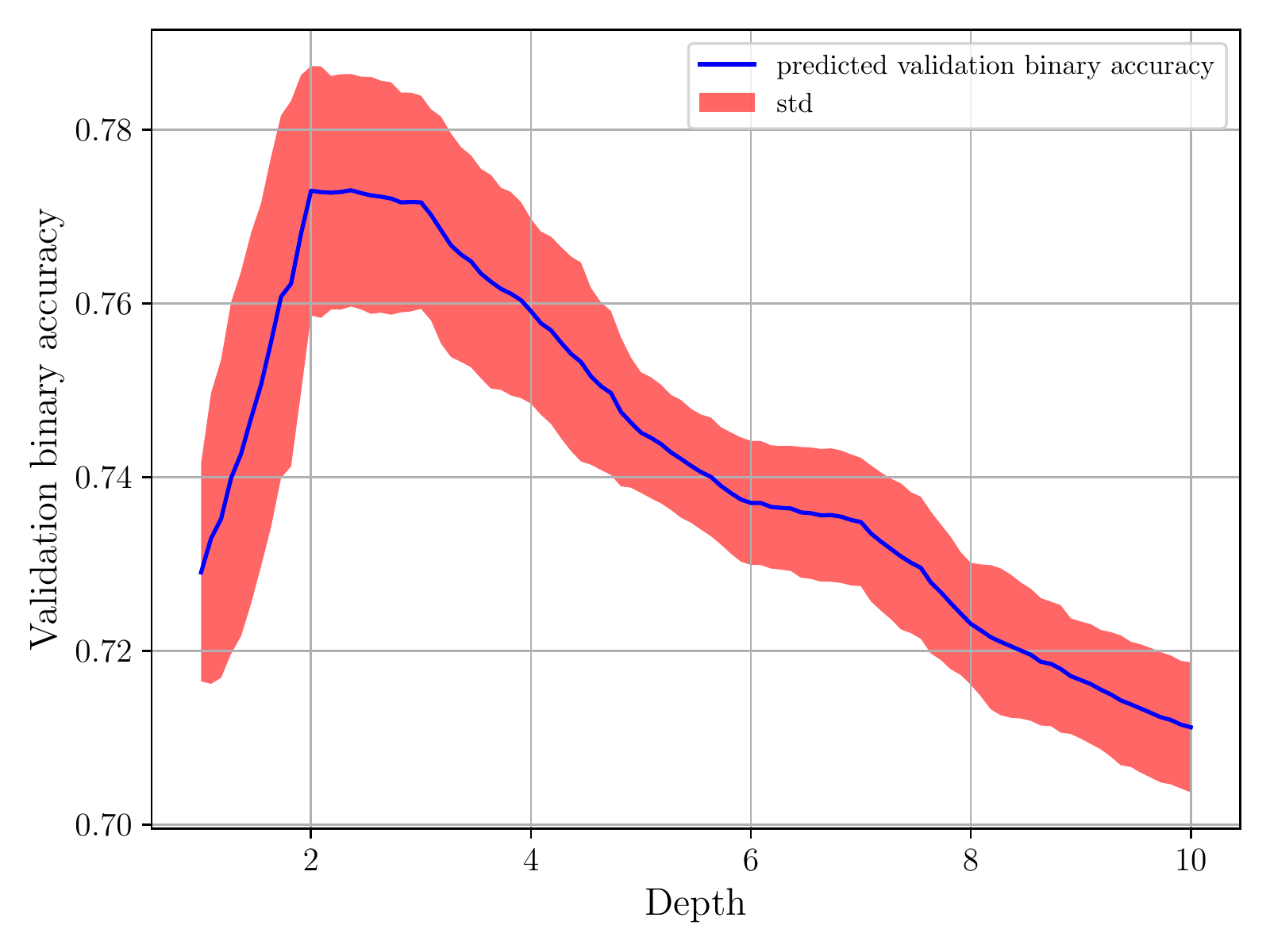}}
\subfloat[]{\includegraphics[width=0.33\textwidth]{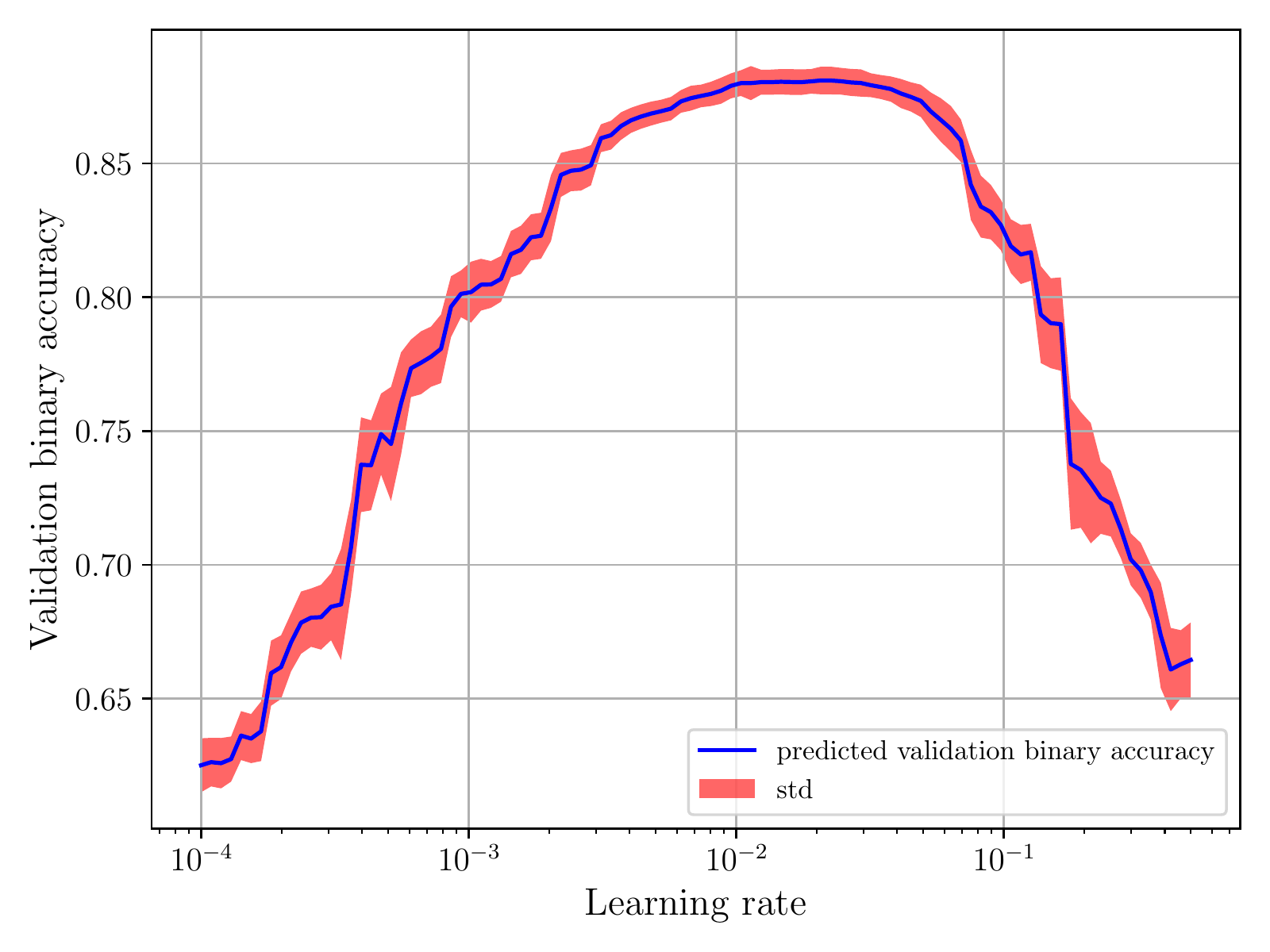}}
\subfloat[]{\includegraphics[width=0.33\textwidth]{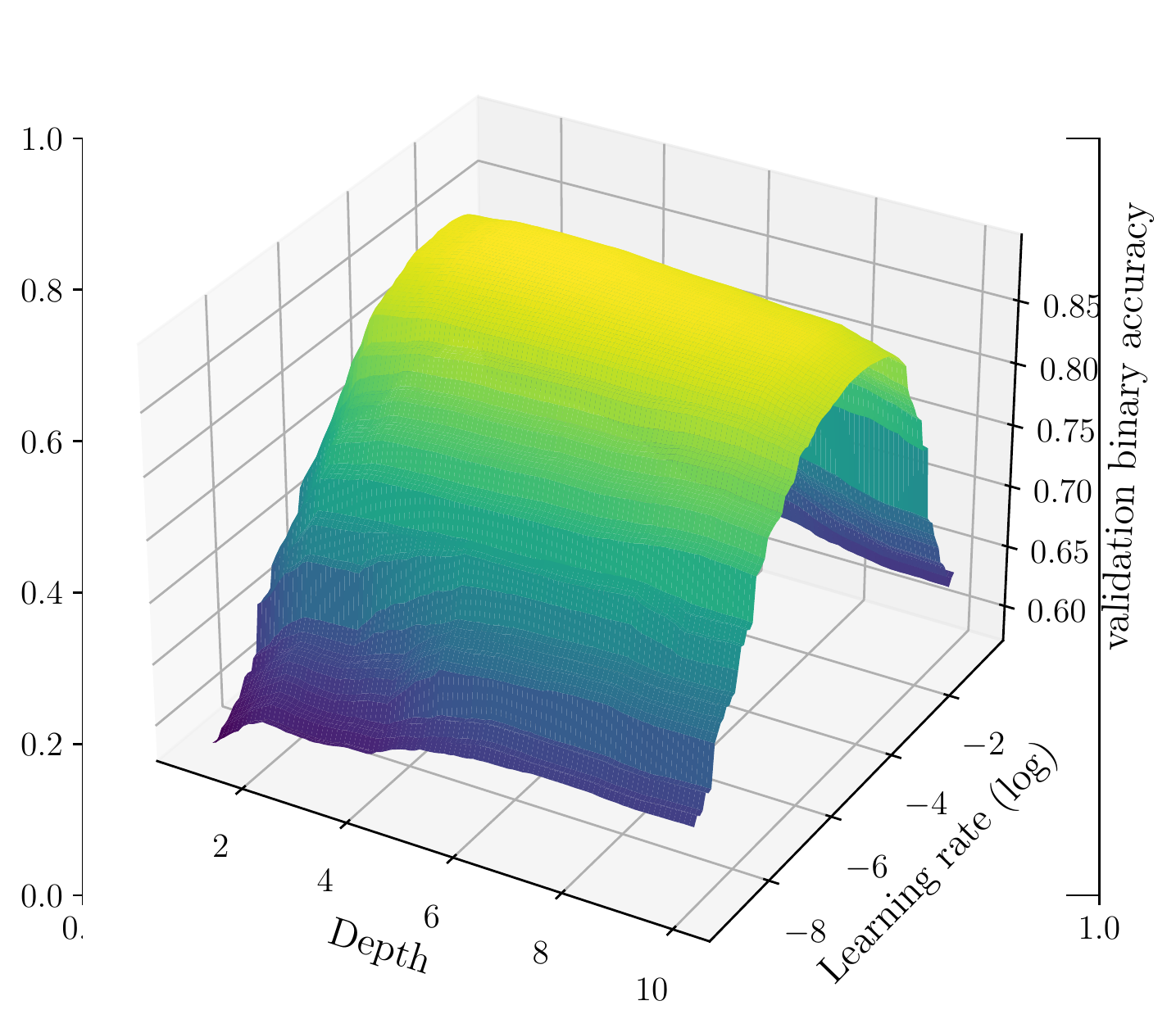}}
\caption{
\small Examples of marginals for a quantum neural network with validation accuracy as performance on the banknote-authentication dataset. The hyperparameters correspond to the learning rate used during training (a), and the number of layers, also known as depth (b), and their combination (c). The hyperparameter values for learning rate are on a log scale. When considered individually, we see for instance that depth and learning rate should not be set too high for better performances. However, when grouped together, the learning rate seems most influential.}
\label{marginalplots}
\end{figure}

Functional ANOVA is based on the concept of a marginal of a hyperparameter, i.e., how a given value for a hyperparameter performs, averaged over all possible combinations of the other hyperparameters’ values. The \emph{marginal performance} $\hat{a}_U(\vtheta_U)$ is described as the average performance of all complete instantiations $\vtheta$ that have the same values for hyperparameters that are in $\vtheta_U$. As an illustration, Fig.~\ref{marginalplots} shows marginals for two hyperparameters of a quantum neural network and their union. As the number of terms to consider for the marginal can be very large, the authors of~\cite{Hutter2014} used tree-based surrogate regression models to calculate efficiently the average performance. Such a model yields predictions $\hat{y}$ for the performance $p$ of arbitrary hyperparameter settings.
\par 
Functional ANOVA determines how much each hyperparameter (and each combination of hyperparameters) contributes to the variance of $\hat{y}$ across the algorithm's hyperparameter space $\vTheta$, denoted $\mathds{V}$.  Intuitively, if the marginal has high variance, the hyperparameter is highly important to the performance measure. Such framework has been used for studying the importance of hyperparameters of common machine learning models such as support vector machines, random forests, Adaboost, and Residual neural networks~\cite{hpimp,resnetfanova}. We refer to~\cite{Hutter2014} for a complete description and introduce the quantum supervised models considered in this study along with the basics of quantum computing. 

\subsection{Supervised learning with Parameterized Quantum Circuits}

\subsubsection{Basics of quantum computing}

In quantum computing, computations are carried out by the manipulation of qubits, similarly to classical computing with bits. A system of $n$ qubits is represented by a $2^n$-dimensional complex vector in the Hilbert space $\mathcal{H}=(\mathbb{C}^2)^{\otimes n}$. This vector describes the state of the system $\ket{\psi} \in \mathcal{H}$ of unit norm $\braket{\psi}=1$. The bra-ket notation is used to describe vectors $\ket{\psi}$, their conjugate transpose $\bra{\psi}$ and inner-products $\braket{\psi}{\psi'}$ in $\mathcal{H}$. Single-qubit computational basis states are given by $\ket{0}=(1,0)^T, \ket{1}=(0,1)^T$, and their tensor products describe general computational basis states, e.g., $\ket{10} = \ket{1}\otimes\ket{0} = (0,0,1,0)$.
\par
The quantum state is modified with unitary operations or gates $U$ acting on $\mathcal{H}$. This computation can be represented by a quantum circuit (see Fig.~\ref{pqcdiagram}). When a gate $U$ acts non-trivially only on a subset $S \subseteq [n]$ of qubits, we denote such operation $U\otimes \mathbbm{1}_{[n]\backslash S}$. In this work, we use, the Hadamard gate $H$, the single-qubit Pauli gates $Z, Y$ and their associated rotations $R_X, R_Y, R_Z$:
\begin{equation}
\begin{gathered}
	H = \frac{1}{\sqrt{2}} \begin{pmatrix} 1 & 1 \\ 1 & -1 \end{pmatrix},
	Z = \begin{pmatrix} 1 & 0 \\ 0 & -1 \end{pmatrix}, R_Z(w) = \exp(-i \frac{w}{2} Z),\\
    Y = \begin{pmatrix} 0 & -i \\ i & 0 \end{pmatrix}, R_Y(w) = \exp(-i \frac{w}{2} Y),
    X = \begin{pmatrix} 0 & 1 \\ 1 & 0 \end{pmatrix}, R_X(w) = \exp(-i \frac{w}{2} X),
    \end{gathered}
\end{equation}

The rotation angles are denoted $w \in\mathbb{R}$ and the 2-qubit controlled-$Z$ gate $\CZ{scale=1.5} = \text{diag}(1,1,1,-1)$ as well as the $\sqrt{\text{iSWAP}}$ given by the matrix 
\begin{equation}
	 \frac{1}{\sqrt{2}} \begin{pmatrix} \sqrt{2} & 0 & 0 & 0 \\ 0 & 1 & i & 0 \\ 0 & i & 1 & 0 \\ 0 & 0 & 0 & \sqrt{2} \end{pmatrix}.
\end{equation}

Measurements are carried out at the end of a quantum circuit to obtain bitstrings. Such measurement operation is described by a Hermitian operator $O$ called an observable. Its spectral decomposition $O=\sum_m \lambda_m P_m$ in terms of eigenvalues $\lambda_m$ and orthogonal projections $P_m$ defines the outcomes of this measurement, according to the Born rule: a measured state $\ket{\psi}$ gives the outcome $\lambda_m$ and gets projected onto the state $P_m \ket{\psi} / \!\ \sqrt{p(m})$ with probability $p(m) = \bra{\psi} P_m \ket{\psi} = \expval{P_m}_\psi$. The expectation value of the observable $O$ with respect to $\ket{\psi}$ is $\mathbb{E}_\psi [O] = \sum_m p(m) \lambda_m = \expval{O}_{\psi}$. We refer to~\cite{nielsenchuang} for more basic concepts of quantum computing, and follow with parameterized quantum circuits.

\subsubsection{Parameterized Quantum Circuits}

A parameterized quantum circuit (also called \emph{ansatz}) can be represented by a quantum circuit with adjustable real-valued parameters $\params$. The latter is then defined by a unitary $U(\params)$ that acts on a fixed $n$-qubit state (e.g., $\ket{0^{\otimes n}}$). The ansatz may be constructed using the formulation of the problem at hand (typically the case in chemistry~\cite{VQE} or optimization~\cite{QAOA}), or with a problem-independent generic construction. The latter are often designated as \emph{hardware-efficient}.
\par
\begin{figure}[tb]
\centering
\[
\Qcircuit @C=.6em @R=.4em {
\lstick{\ket{0}} & \gate{R_X(x_1) }  & \ctrl{1} & \qw & \qw & \control \qw & \gate{R_Y(\theta^1_1) } & \gate{R_Z(\theta^1_2) } \gategroup{1}{3}{4}{8}{.7em}{--}  & \meter \\
\lstick{\ket{0}} & \gate{R_X(x_2) }  & \control \qw & \ctrl{1} & \qw & \qw & \gate{R_Y(\theta^2_1) } & \gate{R_Z(\theta^2_2) } &  \meter \\
\lstick{\ket{0}} & \gate{R_X(x_3) }  & \qw & \control \qw & \ctrl{1} & \qw & \gate{R_Y(\theta^3_1) } & \gate{R_Z(\theta^3_2) } & \meter \\
\lstick{\ket{0}} & \gate{R_X(x_4) }  & \qw & \qw & \control \qw & \ctrl{-3} & \gate{R_Y(\theta^4_1) } & \gate{R_Z(\theta^4_2) }  & \meter \\
}
\]

    \caption{Parameterized quantum circuit architecture example with $4$ qubits and ring connectivity (qubit $1$ is connected to $2$, $2$ to $3$, $3$ to $4$, and $4$ to $1$ makes a ring). The first layer of $R_X$ is the encoding layer $U_\text{enc}$, taking a data instance $x \in \mathbb{R}^4 $ as input. It is followed by the entangling part with Ctrl-$Z$ gates. Finally a variational layer $U_\text{var}$ is applied. Eventually, we do measurements to be converted into predictions for a supervised task. The dashed part can be repeated many times to increase the expressive power of the model.}
    \label{pqcdiagram}
\end{figure}
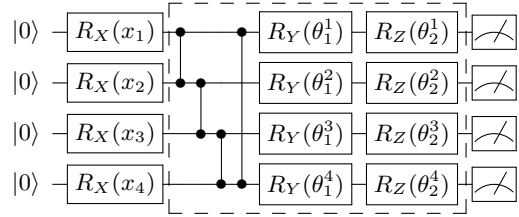

For a machine learning task, this unitary encodes an input data instance $x \in \mathbb{R}^d$ and is parameterized by a trainable vector $\params$. Many designs exist but hardware-efficient parameterized quantum circuits~\cite{hea} with an alternating-layered architecture are often considered in quantum machine learning when no information on the structure of the data is provided. This architecture is depicted in an example presented in Fig.~\ref{pqcdiagram} and essentially consists of an alternation of encoding unitaries $U_\text{enc}$ and variational unitaries $U_\text{var}$. In the example, $U_\text{enc}$ is composed of single-qubit rotations $R_X$, and $U_\text{var}$ of single-qubit rotations $R_z,R_y$ and entangling Ctrl-$Z$ gates, represented as $\CZ{scale=1.5}$ in Fig.~\ref{pqcdiagram}, forming the entangling part of the circuit. Such entangling part denoted $U_\text{ent}$, can be defined by connectivity between qubits. 
\par 
These parameterized quantum circuits are similar to neural networks where the circuit architecture is fixed and the gate parameters are adjusted by a classical optimizer such as gradient descent. They have also been named quantum neural networks. The parameterized layer can be repeated multiple times, which increases its \emph{expressive power} like neural networks~\cite{Sim2019}. The data encoding strategy (such as reusing the encoding layer multiple times in the circuit - a strategy called \emph{data reuploading}) also influences the latter~\cite{datareupload,qfourier}.
\par 
Finally, the user can define the observable(s) and the post-processing method to convert the circuit outputs into a prediction in the case of supervised learning. Commonly, observables based on the single-qubit $Z$ operator are used. When applied on $m \le n$ qubits, the observable is represented by a $2^m-1$ square diagonal matrix with $\{-1,1\}$ values, and is denoted $\mathcal{O} = Z \otimes Z \otimes \cdots \otimes Z $. 
\par 
Having introduced parameterized quantum circuits, we present the hyperparameters of the models, the configuration space, and the experimental setup for our functional ANOVA-based hyperparameter importance study. 

\section{Methods}
\label{methods}

In this section, we describe the network type and its hyperparameters and define the methodology that we follow.

\subsection{Hyperparameters and configuration space}
\label{space}

Many designs have been proposed for parameterized quantum circuits depending on the problem at hand or motivated research questions and contributions. Such propositions can be aggregated and translated into a set of hyperparameters and configuration space for the importance study. As such, we first did an extensive literature review on parameterized quantum circuits for machine learning~\cite{PQC,qsvm,qrobot,qmlbeyondkernel,qrl1,qborn,qsimon,qscreen,qcl,largeqmlgoogle,qnoisy,qrl2,qnat,onchipqnn,qmlreview2,qgan} as well as quantum machine learning software~\cite{qiskit,pennylane,tfq}. This resulted in a list of $10$ hyperparameters, presented in Table~\ref{hparam-table}. We choose them so we balance between having well-known hyperparameters that are expected to be important, and less considered ones in the literature. For instance, many works use Adam~\cite{adam} as the underlying optimizer, and the learning rate should generally be well chosen. On the contrary, the entangling gate used in the parameterized quantum circuit is generally a fixed choice. 
\par
From the literature, we expect data encoding strategy/circuit to be important. We choose two main forms for $U_\text{enc}$. The first one is the hardware-efficient $\bigotimes_{i=1}^{n} R_X(x_i)$. The second takes the following form from~\cite{pennylane,qmlbeyondkernel,qsvm}:
\begin{equation}\label{eq:havlivcek}
    U_\text{enc}(\bm{x}) = U_z(\bm{x})H^{\otimes n}
\end{equation}
\begin{equation}
    U_z(\bm{x}) = \exp(-\mathrm{i}\pi\left[\sum_{i=1}^{n} x_i Z_i + \sum_{\substack{j=1,\\ j>i}}^{n} x_ix_j Z_iZ_j\right]).
\end{equation}

Using data-reuploading~\cite{datareupload} results in a more expressive model~\cite{qfourier}, and this was also demonstrated numerically~\cite{qrl1,datareupload,qrl2}. Finally, pre-processing of the input is also sometimes used in encoding strategies that directly feed input features into Pauli
rotations. It also influences the expressive power of the model~\cite{qfourier}. In this work, we choose a usual activation function $tanh$ commonly used in neural networks. We do so as its range is $[-1,1]$, which is the same as the data features during training after the normalization step.
\par 
The list of hyperparameters we take into account is non-exhaustive. It can be extended at will, at the cost of more software engineering and budget for running experiments. 

\begin{table}[!tb]
\caption{List of hyperparameters considered for hyperparameter importance for quantum neural network, as we named them in our Tensorflow-Quantum code.}
\label{hparam-table}
\begin{center}
\begin{tabularx}{\textwidth}{llX}
\toprule

Hyperparameter & Values &  Description \\
\midrule
Adam learning rate & \makecell[tl]{$[10^{-4} , 0.5]$\\(log)} & The learning rate with which the quantum neural network starts
training. The range was taken from the automated machine learning library Auto-sklearn~\cite{feurerarxiv20a}. We uniformly sample taking the logarithmic scale. \\
batchsize & \makecell[tl]{$16$,\\ $32$,\\ $64$} & Number of samples in one batch of Adam used during training\\
depth  & \makecell[tl]{ $\{1,2,$\\$\cdots, 10\}$ } & Number of variational layers defining the circuit \\
\makecell[tl]{is data_encoding\\hardware efficient}  & \makecell[tl]{True, \\ False} &  Whether we use the hardware-efficient circuit $\bigotimes_{i=1}^{n} R_X(x_i)$ or an IQP circuit defined in Eq.\ref{eq:havlivcek} to encode the input data.\\
use reuploading  & \makecell[tl]{True, \\ False} & Whether the data encoding layer is used before each variational layer or not. \\
have less rotations  & \makecell[tl]{True, \\ False} & If True, only use layers of $R_Y,R_Z$ gates as the variational layer. If False, add a layer of $R_X$ gates. \\
entangler operation  & \makecell[tl]{cz, \\ sqiswap} &  Which entangling gate to use in $U_\text{ent}$\\
map type  & \makecell[tl]{ring,\\ full,\\ pairs} &  The connectivity used for $U_\text{ent}$. The ring connectivity use an entangling gate between consecutive indices $(i, i+1), i \in \{1, \dots, n\}$ of qubits. The full one uses a gate between each pair of indices $(i,j), i < j$. Pairs connects even consecutive indices first, then odd consecutive ones. \\
\makecell[tl]{input activation\\function}  & \makecell[tl]{linear,\\tanh} & Whether to input $tanh(x_i)$ as rotations or just $x_i$. \\
output circuit  & \makecell[tl]{2Z,\\ mZ} & The observable(s) used as output(s) of the circuit. If 2Z, we use all possible pairs of qubit indices defining $Z \otimes Z$. If mZ, the tensor product acts on all qubits. Note we do not use single-qubit $Z$ observables although they are quite often used in the literature. Indeed, they are provably not using the entire circuit when it is shallow. Hence we decided to use $Z \otimes Z$ instead. Also, a single neuron layer with a sigmoid activation function is used as a final decision layer similar to~\cite{qnoisy}. \\

\bottomrule
\end{tabularx}
\end{center}

\end{table}

\subsection{Assessing Hyperparameter Importance}
\label{methodology}

Once the list of hyperparameters and configuration space are decided, we perform the hyperparameter importance analysis with the functional ANOVA framework. Assessing the importance of the hyperparameters boils down to four steps. Firstly, the models are applied to various datasets by sampling various configurations in a hyperparameter optimization process. The performances or metrics of the models are recorded along. The sampled configurations and performances serve as data for functional ANOVA. As functional ANOVA uses internally tree-based surrogate models, namely random forests~\cite{rfcite}, we decided to add an extra step with reference to~\cite{hpimp}. As the second step, we verify the performance of the internal surrogate models. We cross-evaluate them using regression metrics commonly used in surrogate benchmarks~\cite{surrogatehparams}. Surrogates performing badly at this step are then discarded from the importance analysis, as they can deteriorate the quality of the study. Thirdly, the marginal contribution of each hyperparameter over all datasets can be then obtained and used to infer a ranking of their importance. Finally, a verification step similar to~\cite{hpimp} is carried out to confirm the inferred ranking previously obtained. We explain such a procedure in the following section.

\subsection{Verifying Hyperparameter Importance}

\begin{wrapfigure}{r}{0.45\textwidth}
 \vspace{-0.6cm}
  \includegraphics[width=0.42\textwidth]{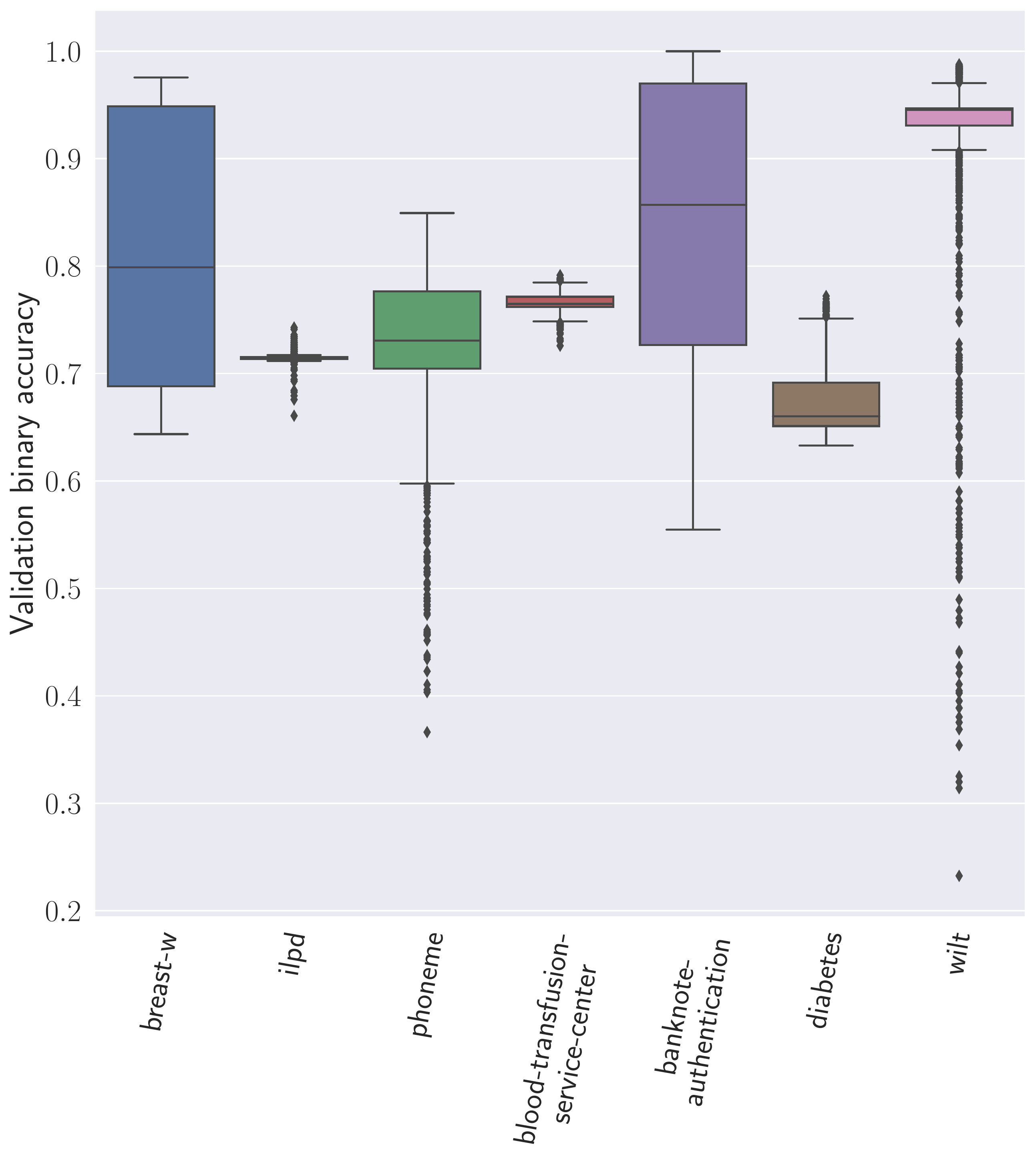}
 \caption{
 \small Performances of $1\,000$ quantim machine learning models defined by different configurations of hyperparameters over each dataset. The metric of interest in the study is the $10$-fold cross-validation accuracy. We take the best-achieved metric per model trained over $100$ epochs.}
 \label{ratios}
 \vspace{-0.6cm}
\end{wrapfigure}
When applying the functional ANOVA framework, an extra verification step is added to confirm the output from a more intuitive notion of hyperparameter importance~\cite{hpimp}. 
It is based on the assumption that hyperparameters that perform badly when fixed to a certain value (while other hyperparameters are optimized), will be important to optimize. 
The authors of~\cite{hpimp} proposed to carry out a costly random search procedure fixing one hyperparameter at a time. 
In order to avoud a bias to the chosen value to which this hyperparameter is fixed, several values are chosen, and the optimization procedure is carried out multiple times.
Formally, for each hyperparameter $\theta_j$ we measure $y^*_{j,f}$ as the result of a random search for maximizing the metric, fixing $\theta_j$ to a given value $f \in F_j, F_j \subseteq \Theta_j$. For categorical $\theta_j$ with domain $\Theta_j$, $F_j=\Theta_j$ is used. For numeric $\theta_j$, the authors of~\cite{hpimp} use a set of $10$ values spread uniformly over $\theta_j$'s range. We then compute $y^*_j = \frac{1}{|F_j|}\sum_{f \in F_j}y^*_{j,f}$, representing the score when not optimizing hyperparameter $\theta_j$, averaged over fixing $\theta_j$ to various values it can take. Hyperparameters with lower value for $y^*_j$ are assumed to be more important, since the performance should deteriorate more when set sub-optimally.
\par 
In our study, we extend this framework to be used on the scale of quantum machine learning models. As quantum simulations can be very expensive, we carry out the verification experiment by using the predictions of the surrogate instead of fitting new quantum models during the verification experiment. The surrogates yield predictions $\hat{y}$ for the performance of arbitrary hyperparameter settings sampled during a random search. Hence, they serve to compute $y^*_{j,f}$. This is also why we assessed the quality of the built-in surrogates as the second step. Poorly-performing surrogates can deteriorate the quality of the random search procedure.

\section{Dataset and inclusion criteria}
\label{setup}

To apply our quantum models and study the importance of the previously introduced hyperparameters, we consider classical datasets. Similarly to~\cite{hpimp}, we use datasets from the OpenML-CC18 benchmark suite~\cite{openmlcc18}. In our study, we consider only the case where the number of qubits available is equal to the number of features, a common setting in the quantum community. As simulating quantum circuits is a costly task, we limit this study to the case where the number of features is less than $20$ after preprocessing\footnote{A $10$-fold cross-validation run in our experiment takes on average $262$ minutes for $100$ epochs with Tensorflow Quantum~\cite{tfq}.}. Our first step was to identify which datasets fit this criterion. We include all datasets from the OpenML-CC18 that have 20 or fewer features after categorical hyperparameters have been one-hot-encoded, and constant features are removed. Afterward, the input variables are also scaled to unit variance as a normalization step. The scaling constants are calculated on the training data and applied to the test data. 
\par 
The final list of datasets is given in Table~\ref{dataset-table}. In total, $7$ datasets fitted the criterion considered in this study. For all of them, we picked the OpenML Task ID giving the $10$-fold cross-validation task. A quantum model is then applied using the latter procedure, with the aforementioned preprocessing steps. 

\begin{table}[!tb]
\caption{List of datasets used in this study. The number of features is obtained after a usual preprocessing used in machine learning methods, such as one-hot-encoding. }
\label{dataset-table}

\begin{center}
\begin{tabularx}{.8\textwidth}{Xrrr}
\toprule

Dataset & \makecell[l]{OpenML\\Task ID} & \makecell[l]{Number of\\features} & \makecell[l]{Number of\\instances}  \\
\midrule
breast-w  & 15                              & $9$  & $699$ \\
diabetes   & 37                             & $8$  & $768$ \\
phoneme  & 9952                             & $5$  & $5\,404$ \\
ilpd & 9971                                 & $11$ & $583$ \\
banknote-authentication  & 10093            & $4$  & $1\,372$ \\
blood-transfusion-service-center & 10\,101  & $4$  & $748$ \\
wilt & 146820                               & $5$  & $4\,839$ \\
\bottomrule
\end{tabularx}
\end{center}

\end{table}

\section{Results}
\label{resexp}

In this section, we present the results obtained using the hyperparameters and the methodology defined in Section \ref{methods} with the datasets described in Section \ref{setup}. First, we show the distribution of performances obtained during a random search where configurations are independently sampled for each dataset. Then we carry out the surrogate verification. Finally, we present the functional ANOVA results in terms of hyperparameter importance with marginal contributions and the random search verification per hyperparameter. 

\subsection{Performance distributions per dataset}

For each dataset, we sampled independently $1\,000$ hyperparameter configurations and run the quantum models for $100$ epochs as budget. As a performance measure, we recorded the best validation accuracy obtained over $100$ epochs. 
Fig.~\ref{ratios} shows the distribution of the $10$-fold cross-validation accuracy obtained per dataset. We observe the impact of hyperparameter optimization by the difference between the least performing and the best model configuration. For instance, on the wilt dataset, the best model gets an accuracy close to $1$, and the least below $0.25$. We can also see that some datasets present a smaller spread of performances. ilpd and blood-transfusion-service-center are in this case. It seems that hyperparameter optimization does not have a real effect, because most hyperparameter configurations give the same result. As such, the surrogates could not differentiate between various configurations. In general, hyperparameter optimization is important for getting high performances per dataset and detecting datasets where the importance study can be applied.

\subsection{Surrogate verification}
Functional ANOVA relies on an internal surrogate model to determine the marginal contribution per hyperparameter. If this surrogate model is not accurate, this can have a severe limitation on the conclusions drawn from functional ANOVA. In this experiment, we verify whether the hyperparameters can explain the performances of the models. Table~\ref{dataset-table-surrogate} shows the performance of the internal surrogate models. We notice low regression scores for the two datasets (less than $0.75$ R2 scores). Hence we remove them from the analysis.

\begin{table}[!tb]
\caption{Performances of the surrogate models built within functional ANOVA over a $10$-fold cross-validation procedure. We present the average coefficient of determination (R2), root mean squared error (RMSE), and Spearman’s rank correlation coefficient (CC). These are common regression metrics for benchmarking surrogate models on hyperparameters~\cite{surrogatehparams}. The surrogates over ilpd and blood-transfusion-service-center obtain low scores (less than $.75$ R2), hence we remove them from the study.}
\label{dataset-table-surrogate}
\begin{center}

\begin{tabularx}{.8\textwidth}{Xrrr}
\toprule

Dataset & R2 score  & ~~~~~RMSE  & ~~~~~~~~~CC \\
\midrule
breast-w & 0.8663 & 0.0436 & 0.9299 \\
diabetes  & 0.7839 & 0.0155 & 0.8456 \\
phoneme  & 0.8649 & 0.0285 & 0.9282 \\
ilpd & 0.1939 & 0.0040 & 0.4530 \\
banknote-authentication  & 0.8579 & 0.0507 & 0.9399 \\
blood-transfusion-service-center & 0.6104 & 0.0056 & 0.8088 \\
wilt & 0.7912 & 0.0515 & 0.8015 \\
\bottomrule
\end{tabularx}

\end{center}
\end{table}

\subsection{Marginal contributions}

\begin{figure}[!ht]
\centering
\subfloat[]{\includegraphics[width=0.4\textwidth]{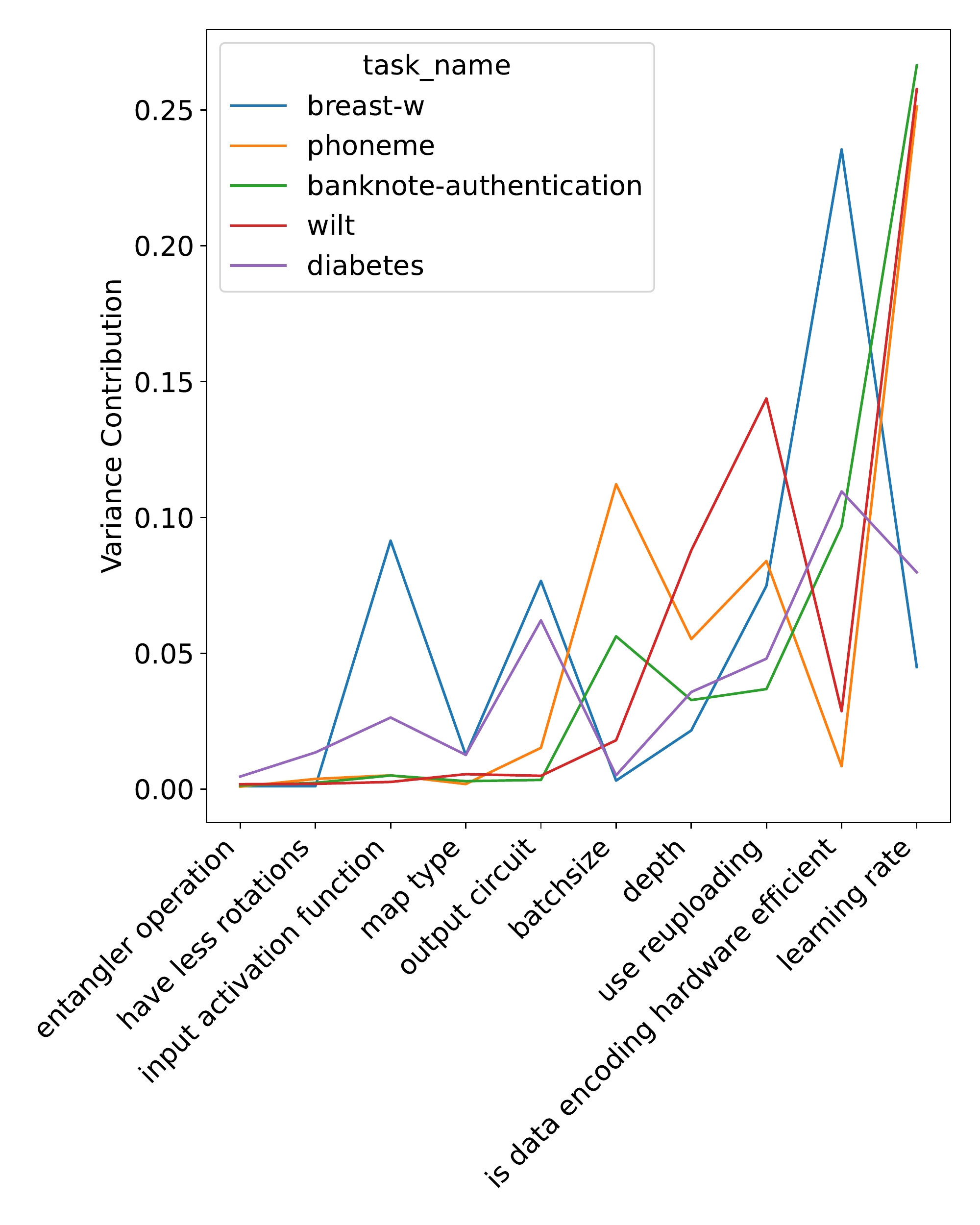}}
\subfloat[]{\includegraphics[width=0.4\textwidth]{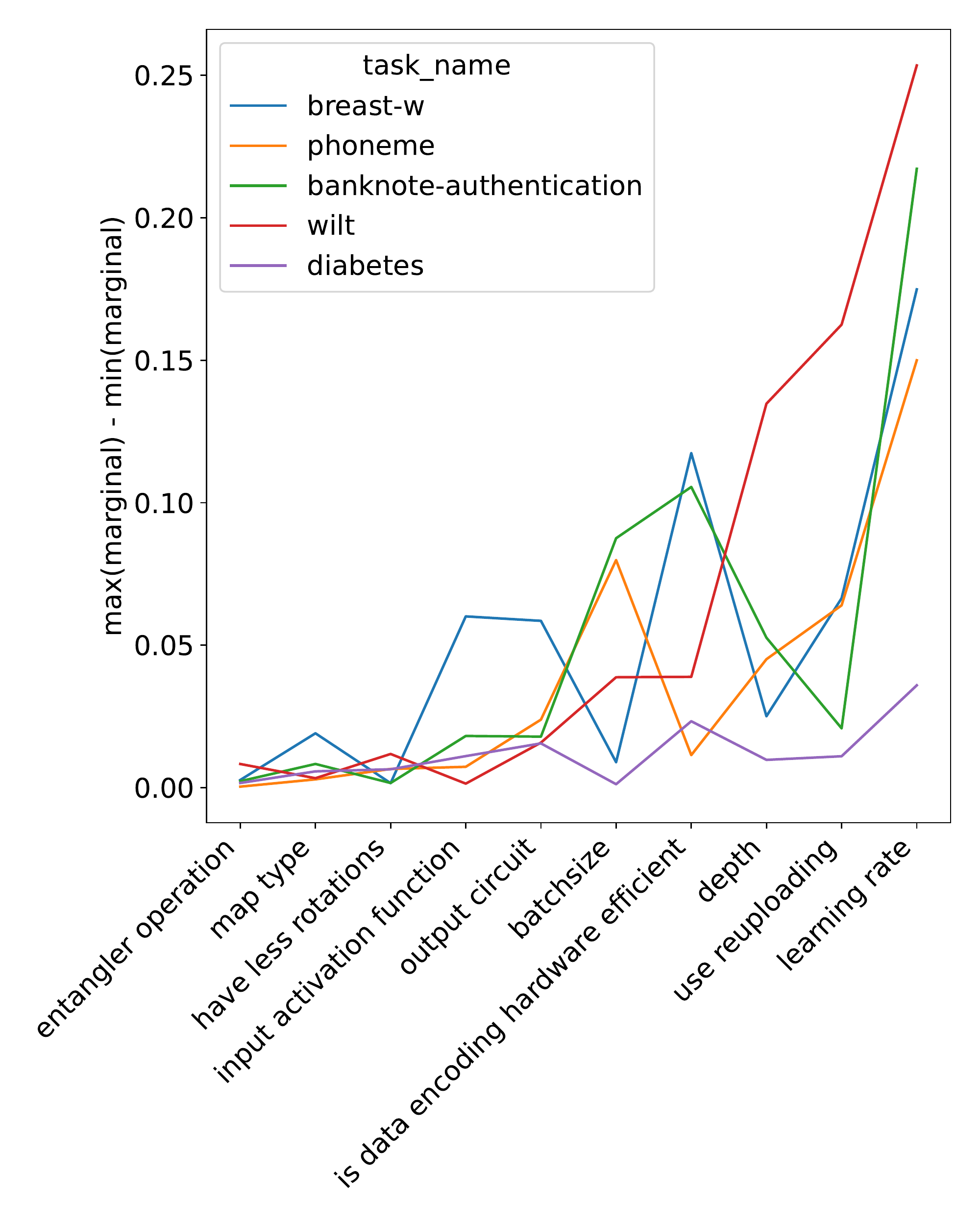}}
\caption{
\small The marginal contributions per dataset are presented as a) the variance contribution  and b) the difference between the minimal and maximal value of the marginal of each hyperparameter. The hyperparameters are sorted from the least to most important using the median. We distinguish from the plot $3$ main levels of importance. 
}
\label{mostimportanthparams}
\end{figure}

For functional ANOVA, we used $128$ trees for the surrogate model. Fig.~\ref{mostimportanthparams}(a,b) shows the marginal contribution of each hyperparameter over the remaining $5$ datasets. We distinguish $3$ main levels of importance. According to these results, the learning rate, depth, and the data encoding circuit and reuploading strategy are critical. These results are in line with our expectations. The entangler gate, connectivity, and whether we use $R_X$ gates in the variational layer are the least important according to functional ANOVA. Hence, our results reveal new insights into these hyperparameters that are not considered in general.

\subsection{Random search verification}

In line with the work of~\cite{hpimp}, we perform an additional verification experiment that verifies whether the outcomes of functional ANOVA are in line with our expectations. 
However, the verification procedure involves an expensive, post-hoc analysis: a random search procedure fixing one hyperparameter at a time. As our quantum simulations are costly, we used the surrogate models fitted on the current dataset considered over the $1\,000$ configurations obtained initially to predict the performances one would obtain when presented with a new configuration.
\par
\begin{wrapfigure}{r}{0.45\textwidth}
  \includegraphics[width=0.45\textwidth]{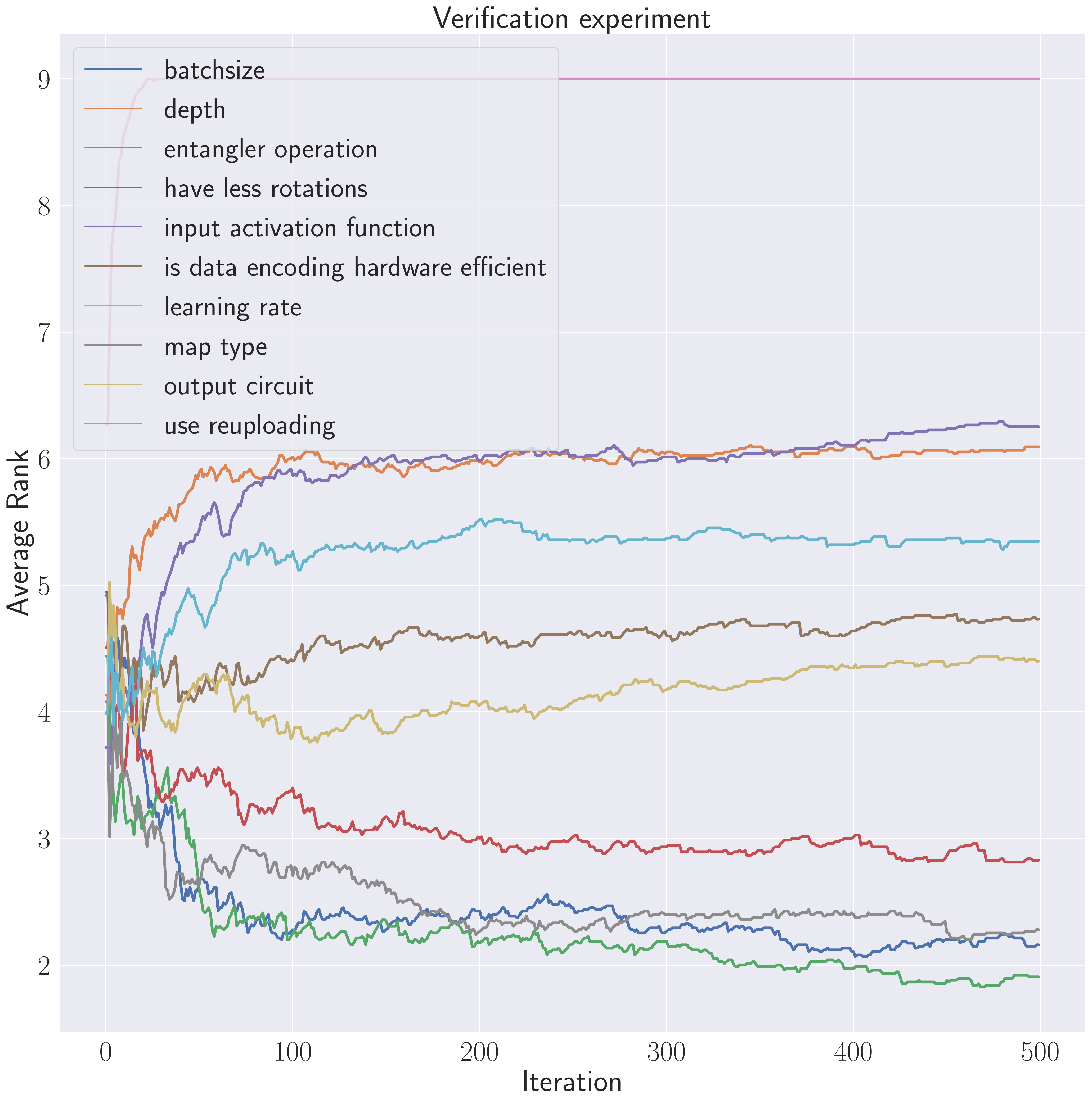}
 \caption{
 \small Verification experiment of the importance of the hyperparameters. A random search procedure up to $500$ iterations excluding one parameter at a time is used. A lower curve means the hyperparameter is deemed less important.}
 \label{fig:verification}
 \vspace{-1.6cm}
\end{wrapfigure}

 
Fig.~\ref{fig:verification} shows the average rank of each run of
random search, labeled with the hyperparameter whose value was
fixed to a default value. A high rank implies poor performance
compared to the other configurations, meaning that tuning this
hyperparameter would have been important. We witness again the $3$ levels of importance, with almost the same order obtained. However, the input\_activation\_function is deemed more important while batchsize is less. 
\par
More simulations with more datasets may be required to validate the importance. However, we retrieve empirically the importance of well-known hyperparameters while considering less important ones. Hence functional ANOVA becomes an interesting tool for quantum machine learning in practice.

\section{Conclusion}
\label{conclusion}

In this work, we study the importance of hyperparameters related to quantum neural networks for classification using the functional ANOVA framework. Our experiments are carried out over OpenML datasets that match the current scale of quantum hardware simulations (i.e., datasets that have at most $20$ features after pre-processing operators have been applied, hence using $20$ qubits). We selected and presented the hyperparameters from an aggregation of quantum computing literature and software. Firstly, hyperparameter optimization highlighted datasets where we observed high differences between configurations. This underlines the importance of hyperparameter optimization for these datasets. There were also datasets that showed little difference. These led us to extend the methodology by adding an additional verification step of the internal surrogate performances. From our results, we distinguished $3$ main levels of importance. On the one hand, Adam's learning rate, depth, and the data encoding strategy are deemed very important, as we expected. On the other hand, the less considered hyperparameters such as the particular choice of the entangling gate and using $3$ rotation types in the variational layer are in the least important group. Hence, our experiment both confirmed expected patterns and revealed new insights for quantum model selection.
\par 
For future work, we plan to further investigate methods from the field of automated machine learning to be applied to quantum neural networks~\cite{brazdil2022metalearning,mohr2022learning,feurerarxiv20a}. 
Indeed, our experiments have shown the importance of hyperparameter optimization, and this should become standard practice and part of the protocols applied within the community. 
We further envision functional ANOVA to be employed in future works related to quantum machine learning and understanding how to apply quantum models in practice. For instance, it would be interesting to consider quantum data, for which quantum machine learning models may have an advantage. Plus, extending hyperparameter importance to techniques for scaling to a large number of features with the number of qubits, such as dimensionality reduction or divide-and-conquer techniques, can be left for future work. Finally, this type of study can also be extended to different noisy hardware and towards algorithm/model selection and design. If we have access to a cluster of different quantum computers, then choosing which hardware works best for machine learning tasks becomes possible. One could also extend our work with meta-learning~\cite{brazdil2022metalearning}, where a model configuration is selected based on meta-features created from dataset features. Such types of studies already exist for parameterized quantum circuits applied to combinatorial optimization~\cite{qalgoselection,Moussa2022,flipinit}.

{\bf Acknowledgements}
CM and VD acknowledge support from TotalEnergies. This work was supported by the Dutch Research Council (NWO/OCW), as part of the Quantum Software Consortium programme (project number 024.003.037). This research is also supported by the project NEASQC funded from the European Union’s Horizon 2020 research and innovation programme (grant agreement No 951821).

\bibliographystyle{splncs04}
\bibliography{references}

\end{document}